\documentclass[a4paper,12pt]{article}
\usepackage{amsmath,amsfonts,amssymb,amsthm,amstext,amscd}
\usepackage{latexsym}
\usepackage{hyperref}
\usepackage{graphicx}
\usepackage{caption}
\usepackage{comment}
\usepackage{cite}
\usepackage{color}
\usepackage{setspace}
\usepackage[mathscr]{eucal}
\usepackage{titletoc}
\usepackage{xcolor}
\usepackage{titlesec}
\newcommand{\mrd}{\mathrm{d}}

\titleformat{\section}
  {\normalfont\large\bfseries}
  {\thesection}
  {1em}
  {}
\titleformat{\subsection}
  {\normalfont\normalsize\bfseries}
  {\thesubsection}
  {1em}
  {}
\definecolor{darkred}{rgb}{0.9,0.05,0.05}
\definecolor{darkblue}{rgb}{0.05,0.05,0.6}
\definecolor{darkgreen}{rgb}{0.05,0.6,0.05}
\newcommand{\txcb}[1]{\textcolor{black}{#1}}% (was darkblue; neutralized to black per request)

\renewcommand*{\eqref}[1]{%
	\begingroup
	\hypersetup{
		linkcolor=darkblue,
		linkbordercolor=darkblue,
	}%
	\textcolor{darkblue}{(\ref{#1})}%
	\endgroup
}
\hypersetup{linkcolor=black,citecolor=darkgreen,urlcolor=black,colorlinks=true}
\begin{document}

\setlength{\skip\footins}{0.8cm}

\begin{titlepage}

\vspace{0.5cm}

\newcommand\blfootnote[1]{%
	\begingroup
	\renewcommand\thefootnote{}\footnote{#1}%
	\addtocounter{footnote}{-1}%
	\endgroup
}

\begin{center}
\renewcommand{\baselinestretch}{1.5}  %Line spacing
\setstretch{1.5}

{\fontsize{17pt}{12pt}\bf{\txcb{Dynamical Completion of Coupling-Charge Thermodynamics}}}
 
\vspace{9mm}
\renewcommand{\baselinestretch}{1}  %Line spacing
\setstretch{1}

\centerline{  {Kamal Hajian$^{\ast\dagger}$\footnote{kamal.hajian@uni-oldenburg.de}} and  {Bayram  Tekin$^{\ddagger}$\footnote{bayram.tekin@bilkent.edu.tr}} }
\vspace{4mm}
\normalsize
$^\ast$\textit{Department of Physics, Middle East Technical University, 06800, Ankara, Turkey}\\
$^\dagger$\textit{Institute of Physics, University of Oldenburg, D-26111 Oldenburg, Germany}\\
$^\ddagger$\textit{Department of Physics, Bilkent University, 06800, Ankara, Turkey}\\
\vspace{5mm}

\begin{abstract}
The scalar-gauge pair formulation promotes gravitational coupling constants to conserved charges, clarifying their role in extended black hole thermodynamics. However, its original incarnation restricts these auxiliary scalar fields to rigid, non-dynamical configurations. In this paper, we introduce a local, dynamical completion of this framework. By constructing the kinetic sector from a defect, which is the difference between the top-form field strength and the associated gravitational Lagrangian density, we allow the scalar to propagate as a genuine massive degree of freedom. Crucially, this local sector acts as a charge-preserving massive scalar backreaction, ensuring the thermodynamic coupling remains a conserved integration constant rather than a locally varying field. When evaluated on constant-coupling configurations, the auxiliary stress-energy contributions vanish entirely, perfectly preserving the original black hole solutions, the first law, and the Smarr relation. We illustrate this construction using the cosmological constant and the Gauss-Bonnet coupling.
\end{abstract}
\end{center}

\let\newpage\relax

\end{titlepage}

\section{Introduction}

The first law and the integrated Smarr relation are cornerstones of black hole thermodynamics in General Relativity \cite{Smarr:1972kt,Bardeen:1973gd,Bekenstein:1973ft,Hawking:1976rt}.  In extended gravity theories with several dimensionful parameters, however, the standard Smarr relation is incomplete unless those parameters are allowed to have (thermodynamic) variations.  This is familiar in the field of  ``black hole chemistry," where the cosmological constant is treated as a pressure, and in higher-curvature gravities, where the Lovelock or other higher-derivative couplings have conjugate potentials \cite{Kastor:2010gq,Gunasekaran:2012dq,Frassino:2014pha,Einstein:1917}. 

The point of the recently advocated scalar-gauge pair construction is to give this extended thermodynamic phase space a covariant field-theoretic origin.  For a theory whose Lagrangian \(D\)-form is split as
\begin{equation}\label{original-Lagrangian}
        \mathbf L=\mathbf L_0-\sum_i \alpha_i\mathbf L_i,
\end{equation}
one introduces, for each coupling, a scalar \(\alpha_i(x)\) and a \((D-1)\)-form gauge field \(\boldsymbol{\mathrm{A}}_i\), with \(\boldsymbol{\mathrm{F}}_i=\mrd\boldsymbol{\mathrm{A}}_i\)\cite{Hajian:2023bhq} (see Refs. \cite{1,Hajian:2021hje,Meessen:2022hcg} for precursors of this formulation).  The extended Lagrangian is 
\begin{equation}\label{topological-pair-lagrangian}
        \widetilde{\mathbf L}
        =\mathbf L_0+
        \sum_i\alpha_i(x)(\boldsymbol{\mathrm{F}}_i-\mathbf L_i).
\end{equation}
Here \(\mathbf L_i\) is simply the \(D\)-form density multiplying the coupling one wishes to promote; it need not be an interaction term, and in pure gravity, it may contain the kinetic curvature density itself.  This distinction is important because promoting an overall factor multiplying a kinetic term can change the constraint structure.  We therefore regard the admissibility condition discussed in Ref.~\cite{Cheong:2026tkv} as part of the definition of the allowed split.  With the volume form \(\boldsymbol\epsilon\), we write \(\mathbf L_i=\mathcal L_i\boldsymbol\epsilon\) and define the scalar top-form component by
\begin{equation}\label{F-convention}
        \boldsymbol{\mathrm{F}}_i=F_i\boldsymbol\epsilon ,
\end{equation}
so that \(F_i\) is simply the component of the top-form \(\boldsymbol{\mathrm{F}}_i\) along the volume form. 
The equations of motion of \eqref{topological-pair-lagrangian} enforce \(\mrd\alpha_i=0\) and \(\boldsymbol{\mathrm{F}}_i=\mathbf L_i\). 
As for recovering \eqref{original-Lagrangian} from \eqref{topological-pair-lagrangian}, one first uses the fact that the field equations demand $\alpha_i$ to be a constant, which, when used in the Lagrangian, yields $\alpha_i \boldsymbol{\mathrm{F}}_i= \mrd(\alpha_i \boldsymbol{\mathrm{A}}_i)$, i.e., a boundary term.
Thus, the original theory is recovered, while the constant value of \(\alpha_i\) becomes the charge conjugate to the form-field potential \cite{1,Hajian:2021hje,Hajian:2023bhq}. Specifically, the gauge symmetry $\boldsymbol{\mathrm{A}}_i \to \boldsymbol{\mathrm{A}}_i + \mrd\boldsymbol{\lambda}_i$ promotes each coupling $\alpha_i$ to be the conserved charge associated with the global part $\mrd \boldsymbol{\lambda}_i=0$. The black hole thermodynamic conjugate variables to these charges naturally emerge as the horizon ``electric" potentials of the corresponding gauge fields. This formulation has been successful in describing the quasi-local and thermodynamic properties of stationary black hole solutions \cite{Hajian:2025hxf}.

The aim of this paper is to ask whether the coupling-charge formulation can be enlarged into a genuinely local theory without losing the black hole sector that gave the construction its thermodynamic meaning.  Our answer is conservative.  The additional dynamics are arranged so that, in the original constant-coupling configurations, the new sector vanishes, and the constant-coupling black hole solutions, the first law, and the Smarr relation are recovered unchanged.  Away from this sector, the new scalar field propagates as an ordinary massive degree of freedom and can backreact on the geometry.  It should not, however, be interpreted as a literal spacetime-dependent replacement of the original coupling in the gravitational field equations.  The thermodynamic coupling remains a conserved integration constant, while the local scalar describes charge-preserving massive excitations around it.  In this sense, the construction is related in spirit to both four-form descriptions of integration constants and scalar-field models \cite{Aurilia:1980xj,Duff:1980qv,Hawking:1984hk,Bousso:2000xa,BrownTeitelboim:1987,Henneaux:1984ji,Henneaux:1989zc,Teitelboim:1985dp}, but it is not, by itself, a dark-energy or quintessence model \cite{Ratra:1988jz, Zlatev:1998tr}; such an interpretation would require a separate cosmological analysis.

\section{\txcb{Defect kinetic completion and effective dynamics}}

The  pair Lagrangian \eqref{topological-pair-lagrangian} has no local degrees of freedom in the auxiliary sector.  To add dynamics while preserving the constant-coupling solution space, we use only the defect 
\begin{equation} 
\boldsymbol\Delta_i :=\boldsymbol{\mathrm{F}}_i-\mathbf L_i.
\end{equation}
The proposed dynamical completion is
\begin{equation}\label{L_dyn}
\widetilde{\mathbf L}_{\rm dyn}
=\mathbf L_0+
\sum_i\alpha_i(x)\boldsymbol\Delta_i
-\frac12\sum_i\kappa_i\,\mrd\alpha_i\wedge\star\mrd\alpha_i
-\frac12\sum_i\hat\kappa_i\,\boldsymbol\Delta_i\wedge\star\boldsymbol\Delta_i ,
\end{equation}
where the constants \(\kappa_i\) and \(\hat\kappa_i\) are not original couplings of gravitational theory; they are new parameters controlling the kinetic completion.  The action remains invariant under the form-field gauge symmetry \(\boldsymbol{\mathrm{A}}_i\to\boldsymbol{\mathrm{A}}_i+\mrd\boldsymbol\lambda_i\).

The reason for using \(\boldsymbol\Delta_i\), rather than \(\boldsymbol{\mathrm{F}}_i\), is simple.  If one added a conventional term \(-\frac12\boldsymbol{\mathrm{F}}_i\wedge\star\boldsymbol{\mathrm{F}}_i\), the gauge equation would relate \(F_i\) directly to \(\alpha_i\) up to an integration constant.  On a constant-\(\alpha_i\) background, this would force \(F_i\) to be constant.  But the auxiliary pair requires \(F_i=\mathcal L_i(x)\), and a generic density \(\mathcal L_i\) is not constant on a black hole spacetime (see such examples in Ref. \cite{Hajian:2025hxf}).  The defect construction avoids this obstruction because \(\boldsymbol\Delta_i=0\) is solved by \(F_i=\mathcal L_i(x)\), even when \(\mathcal L_i\) varies over the spacetime.

It is useful to express \eqref{L_dyn} as a scalar density in order to study the corresponding field equations.  With the convention
\(\boldsymbol\Delta_i=\Delta_i\boldsymbol\epsilon\), one has
\begin{equation}
\boldsymbol\Delta_i\wedge\star\boldsymbol\Delta_i=-\Delta_i^2\boldsymbol\epsilon, \qquad \Delta_i=F_i-\mathcal{L}_i,
\end{equation}
and therefore
\begin{equation}\label{Scalar Lagrangian}
\widetilde{\mathcal L}_{\rm dyn}
=
\mathcal L_0
+\sum_i\alpha_i\Delta_i
-\frac12\sum_i\kappa_i\nabla_\mu\alpha_i\nabla^\mu\alpha_i
+\frac12\sum_i\hat\kappa_i\Delta_i^2 .
\end{equation}
The last term has the opposite sign from an ordinary scalar potential before
the top-form is eliminated.  This does not signal a ghost.  The fundamental
field \(\boldsymbol{\mathrm A}_i\) is a \((D-1)\)-form and carries no local
propagating degree of freedom in \(D\) dimensions.  The scalar component
\(\Delta_i\) is therefore not an independent propagating scalar; it is fixed
locally by the form-field equation, up to an integration constant.

Varying \eqref{Scalar Lagrangian} with respect to \(\alpha_i\) gives
\begin{equation}\label{alpha-eq}
        \Delta_i+\kappa_i\Box\alpha_i=0 .
\end{equation}
Varying \eqref{Scalar Lagrangian} with respect to the \((D-1)\)-form potential
\(\boldsymbol{\mathrm A}_i\) gives
\begin{equation}\label{A-eq}
        \nabla_\mu\alpha_i+\hat\kappa_i\nabla_\mu\Delta_i=0,
\end{equation}
which integrates locally to
\begin{equation}\label{Sol 1}
        \alpha_i+\hat\kappa_i\Delta_i=\bar\alpha_i ,
\end{equation}
where \(\bar\alpha_i\) is constant on each connected component. 
Substituting this relation into \eqref{alpha-eq} gives
\begin{equation}\label{wave}
        (\Box-m_i^2)(\alpha_i-\bar\alpha_i)=0,
        \qquad
        m_i^2=\frac{1}{\kappa_i\hat\kappa_i} .
\end{equation}
Thus the local field \(\alpha_i(x)\) is a massive scalar fluctuation about the
thermodynamic coupling charge \(\bar\alpha_i\).  Note that, although
\(\kappa_i\) and \(\hat\kappa_i\) are independent parameters, their product
must have mass dimension \(-2\), so that \(m_i^2\) has the usual mass
dimension.

For the variation of the Lagrangian \eqref{Scalar Lagrangian} with respect to the metric, let
\begin{equation}
\mathcal E^{(0)}_{\mu\nu}
:=
\frac{2}{\sqrt{-g}}
\frac{\delta(\sqrt{-g}\mathcal L_0)}{\delta g^{\mu\nu}},
\qquad
\mathcal E^{(i)}_{\mu\nu}
:=
\frac{2}{\sqrt{-g}}
\frac{\delta(\sqrt{-g}\mathcal L_i)}{\delta g^{\mu\nu}} .
\end{equation}
Then, using the relation
\begin{align}\label{identity}
\delta_g \left[ \sqrt{-g} \left( \alpha_i\Delta_i + \frac{1}{2} \hat{\kappa}_i \Delta_i^2 \right) \right] &= (\alpha_i + \hat{\kappa}_i \Delta_i) \delta_g(\sqrt{-g} \Delta_i) - \frac{1}{2} \hat{\kappa}_i \Delta_i^2 \delta_g \sqrt{-g}\\
& = -(\alpha_i + \hat{\kappa}_i \Delta_i) \delta_g(\sqrt{-g} \mathcal{L}_i) - \frac{1}{2} \hat{\kappa}_i \Delta_i^2 \delta_g \sqrt{-g},
\end{align}
we find the field equation
\begin{equation}\label{g-eq}
\mathcal{E}_{\mu\nu}^{(0)} - \sum_{i} \left( (\alpha_i + \hat{\kappa}_i \Delta_i) \mathcal{E}_{\mu\nu}^{(i)} + \kappa_i ( \nabla_\mu \alpha_i \nabla_\nu \alpha_i - \frac{1}{2} g_{\mu\nu} (\nabla \alpha_i)^2 ) - \frac{1}{2} \hat{\kappa}_i \Delta_i^2 g_{\mu\nu} \right) = 0.
\end{equation}
Using also \eqref{Sol 1}, the metric equation becomes
\begin{equation}\label{reduced metric eq}
\mathcal E^{(0)}_{\mu\nu}
-\sum_i\bar\alpha_i\mathcal E^{(i)}_{\mu\nu}
-\sum_i\kappa_i
\left(
\nabla_\mu\alpha_i\nabla_\nu\alpha_i
-\frac12g_{\mu\nu}(\nabla\alpha_i)^2
-\frac12m_i^2(\alpha_i-\bar\alpha_i)^2g_{\mu\nu}
\right)
=0 .
\end{equation}
Equivalently, the original sector is governed by the constant couplings
\(\bar\alpha_i\), while the new sector supplies the stress tensor of massive
scalars with potentials
\begin{equation}\label{scalar-potential}
        V_i(\alpha_i)
        =
        \frac{1}{2\hat\kappa_i}
        (\alpha_i-\bar\alpha_i)^2 .
\end{equation}

We note that the scalar stress tensor is conserved by virtue of the scalar equation.  Focusing on one pair for simplicity,
\begin{equation}\label{scalar-stress}
T^{(\alpha)}_{\mu\nu}
=
\kappa
\left(
\nabla_\mu\alpha\nabla_\nu\alpha
-\frac12g_{\mu\nu}(\nabla\alpha)^2
-\frac12m^2(\alpha-\bar\alpha)^2g_{\mu\nu}
\right),
\end{equation}
and hence
\begin{equation}\label{stress-conservation}
\nabla^\mu T^{(\alpha)}_{\mu\nu}
=
\kappa\nabla_\nu\alpha
\left(
\Box\alpha-m^2(\alpha-\bar\alpha)
\right)
=0 .
\end{equation}
This is the consistency condition required by the generalized Bianchi
identity for the original geometric sector with constant coefficients
\(\bar\alpha_i\).  Indeed, for the metric geometric densities considered here, the corresponding Euler tensors satisfy the usual Noether identity and are identically divergence-free.
\begin{equation}
        \nabla^\mu
        \left(
        \mathcal E^{(0)}_{\mu\nu}
        -
        \sum_i\bar\alpha_i\mathcal E^{(i)}_{\mu\nu}
        \right)
        =
        0 .
\end{equation}
If a spacetime-dependent $\alpha_i(x)$ were to multiply $\mathcal E^{(i)}_{\mu\nu}$ directly, extra terms proportional to $\nabla_\mu\alpha_i$ would appear.  The form-field equation prevents
this by locking the geometric coefficient to the conserved charge
\(\bar\alpha_i\).

For non-metric matter fields \(\Psi\), the same point appears already at the
level of the variational principle.  Since the defect contains
\(-\mathcal L_i\), one has, before using the form-field equation,
\begin{equation}\label{matter-offshell}
        \delta_\Psi\widetilde{\mathcal L}_{\rm dyn}=\delta_\Psi\mathcal L_0-\sum_i(\alpha_i+\hat\kappa_i\Delta_i)\delta_\Psi\mathcal L_i,
\end{equation}
in which we have used the relation
\begin{equation}
\delta_\Psi(\alpha_i\Delta_i+\frac{1}{2} \hat{\kappa}_i \Delta_i^2)=(\alpha_i+\hat\kappa_i\Delta_i)\delta_\Psi\Delta_i=-(\alpha_i+\hat\kappa_i\Delta_i)\delta_\Psi\mathcal{L}_i.
\end{equation}
After imposing \eqref{Sol 1}, however, the coefficient
is constant and the matter equation reduces to the original one with
\(\bar\alpha_i\):
\begin{equation}\label{matter-onshell}
        \delta_\Psi\mathcal L_0
        -
        \sum_i\bar\alpha_i\,\delta_\Psi\mathcal L_i
        =
        0 .
\end{equation}
Local variations of \(\alpha_i(x)\) therefore do not change the matter
couplings in the original equations.  They enter the metric equation only
through the massive scalar stress tensor. In this sense, the construction gives a massive scalar completion of coupling-charge thermodynamics.

It is easy to see that \(\alpha_i=\bar\alpha_i\) is a solution to the coupled field equations \eqref{alpha-eq}, \eqref{A-eq}, \eqref{g-eq}, and \eqref{matter-onshell} with \(\Delta_i=0\) and the original field equations. In other words, the constant-coupling backgrounds are recovered at the minimum of the scalar field potential, and therefore, the original constant-coupling sector is an exact sector of the new theory.  In this case, the metric and matter equations reduce exactly to those of the original theory with constant coupling.  When \(\alpha_i\) is excited, the new sector backreacts, but, as we now show, the backreaction is that of a massive scalar rather than a literal replacement \(\alpha_i\to\alpha_i(x)\) in the original field equations. The mechanism is the form-field equation, which ensures that the combination multiplying the original density remains constant.

\section{\txcb{Boundary conditions, charges, and degrees of freedom}}\label{sec:dof}

{\color{black}The enlarged theory should preserve the asymptotic coupling charge carried by \(\bar\alpha_i\).  This requires the massive scalar excitation to be localized or normalizable.  In an asymptotically flat region, the regular massive mode behaves as derived in Appendix A.  
\begin{equation}\label{falloff-flat}
        \alpha_i(r)-\bar\alpha_i
        \sim\frac{e^{-m_ir}}{r^{(D-2)/2}} .
\end{equation}
In asymptotically AdS spacetime of radius \(\ell\), the normalizable branch is
\begin{equation}\label{falloff-ads}
        \alpha_i(r)-\bar\alpha_i\sim r^{-\delta_+},
        \qquad
        \delta_+=\frac{D-1}{2}+
        \sqrt{\frac{(D-1)^2}{4}+m_i^2\ell^2} .
\end{equation}
These falloffs imply \(\Delta_i=(\bar\alpha_i-\alpha_i)/\hat\kappa_i\to0\) at the boundary. Thus, the defect sector does not shift the asymptotic value of the coupling charge.

The covariant phase-space, or Lee-Wald/Iyer-Wald, formulation \cite{Lee:1990gr, Wald:1993nt, Iyer:1994ys,Wald:1999wa,Ashtekar:1987hia,Ashtekar:1990gc,Barnich:2001jy,Crnkovic:1987at} (see reviews in \cite{Hajian:2015eha,Hajian:2015xlp} and applications in \cite{Hajian:2016kxx,Ghodrati:2016vvf}) is also sharpened by \eqref{Sol 1}.  The variation with respect to the form field produces the auxiliary symplectic potential
\begin{equation}
        \widetilde\Theta_{\boldsymbol A}
        =\sum_i(\alpha_i+\hat\kappa_i\Delta_i)\,\delta\boldsymbol{\mathrm{A}}_i .
\end{equation}
With the same normalization as in the auxiliary pair formalism, 
\begin{equation}
        \widetilde\Theta_{\boldsymbol A}\big|_{\rm on\ shell}
        =\sum_i\bar\alpha_i\,\delta\boldsymbol{\mathrm{A}}_i,
\end{equation}
and the integrable charge associated with the global form-field symmetry is identified with \cite{Hajian:2023bhq}
\begin{equation}\label{coupling charge}
        H_{\boldsymbol\lambda_i}=\bar\alpha_i.
\end{equation}
The local field \(\alpha_i(x)\) is not itself the thermodynamic charge once the kinetic completion is turned on.  A fully explicit surface-charge derivation requires specifying the allowed gauge parameters, the boundary conditions on \(\boldsymbol{\mathrm{A}}_i\), and the representative of the covariant phase-space potential.  For the present purpose, the important point is the on-shell cancellation: the boundary term depends on \(\alpha_i+\hat\kappa_i\Delta_i\), and this combination is exactly \(\bar\alpha_i\).

We next count the local degrees of freedom.  For one pair, the scalar density is
\begin{equation}
\mathcal L_{\rm pair}
=\alpha\Delta-\frac12\kappa(\nabla\alpha)^2+\frac12\hat\kappa\Delta^2 .
\end{equation}
The top-form component \(\Delta\) is algebraic.  Solving the equation gives
\begin{equation}
        \Delta=\frac{\bar\alpha-\alpha}{\hat\kappa},
\end{equation}
and one obtains the reduced bulk
representative
\begin{equation}
        \mathcal L_{\rm eff}
        =
        \mathcal L_0-\bar\alpha\mathcal L
        -\frac12\kappa\nabla_\mu\alpha\nabla^\mu\alpha
        -\frac{1}{2\hat\kappa}(\alpha-\bar\alpha)^2 .
\end{equation}
A \((D-1)\)-form has no local propagating degrees of freedom in \(D\) dimensions, so each pair adds exactly one scalar mode.  The Hamiltonian density of this mode is
\begin{equation}
        \mathcal H_{\rm eff}
        =\frac{\pi_\alpha^2}{2\kappa}
        +\frac{\kappa}{2}(\nabla_a\alpha)^2
        +\frac{1}{2\hat\kappa}(\alpha-\bar\alpha)^2 .
\end{equation}
It is bounded below provided
\begin{equation}
        \kappa>0,
        \qquad
        \hat\kappa>0 .
\end{equation}
These conditions are stronger than \(\kappa\hat\kappa>0\), which only guarantees \(m^2>0\).  With these signs, the apparent wrong-sign \(+\frac12\hat\kappa\Delta^2\) in \eqref{Scalar Lagrangian} is harmless because \(\Delta\) is auxiliary; the physical scalar has the standard kinetic sign and a positive quadratic potential.

The recent constraint analysis of Ref.~\cite{Cheong:2026tkv} is relevant here.  It warns that not every formal promotion of a coupling is dynamically innocuous, especially if one promotes an overall coefficient multiplying kinetic terms.  Our construction should therefore be read with the same restriction already imposed in the topological split: only admissible couplings \(\alpha_i\) are promoted, and the kinetic completion is added after that split has been chosen.  Under this assumption, the new local content is exactly the massive scalar described above; the top-form still carries no propagating local mode.
}

\section{\txcb{Examples: cosmological and Gauss-Bonnet couplings}}

{\color{black}\subsection{The cosmological term}

The cosmological term serves as the simplest example because the $\mathcal{L}_\Lambda$ is a constant, which can be scaled to $1$ simply by redefining $\Lambda$ (for example, with the General Relativity Lagrangian $\mathcal{L}_{\text{GR}}=\frac{1}{16\pi G}(R-2\Lambda_0)$, this is done by letting $\Lambda=\frac{\Lambda_0}{8\pi G}$).   Let us consider a generic theory of gravity where a spacetime-dependent $\Lambda(x)$ is coupled to a $(D-1)$-form field, $\boldsymbol{\mathrm{A}}_\Lambda$, and define the scalar top-form component by
\begin{equation}
        \mrd\boldsymbol{\mathrm A}_\Lambda=F_\Lambda\boldsymbol\epsilon,
        \qquad
        \Delta_\Lambda=F_\Lambda-1 .
\end{equation}
The pair contribution to the scalar density is
\begin{equation}
\mathcal L_{\Lambda,{\rm pair}}
=\Lambda(F_\Lambda-1)
-\frac12\kappa_\Lambda\nabla_\mu\Lambda\nabla^\mu\Lambda
+\frac12\hat\kappa_\Lambda(F_\Lambda-1)^2 .
\end{equation}
The equations are
\begin{align}\label{Lambda EOMs}
&(F_\Lambda-1)+\kappa_\Lambda\Box\Lambda=0,
\qquad
\nabla_\mu\Lambda+\hat\kappa_\Lambda\nabla_\mu F_\Lambda=0,
\nonumber\\
&\mathcal E^{(0)}_{\mu\nu}
-\left[\Lambda+
\hat\kappa_\Lambda(F_\Lambda-1)\right]\mathcal E^{(\Lambda)}_{\mu\nu}
-\kappa_\Lambda\left(
\nabla_\mu\Lambda\nabla_\nu\Lambda
-\frac12g_{\mu\nu}(\nabla\Lambda)^2
\right)
+\frac12\hat\kappa_\Lambda(F_\Lambda-1)^2g_{\mu\nu}=0 .
\end{align}
Integrating the gauge equation gives
\begin{equation}
        \Lambda+
        \hat\kappa_\Lambda(F_\Lambda-1)=\bar\Lambda .
\end{equation}
Thus
\begin{equation}
        (\Box-m_\Lambda^2)(\Lambda-\bar\Lambda)=0,
        \qquad
        m_\Lambda^2=\frac{1}{\kappa_\Lambda\hat\kappa_\Lambda},
\end{equation}
and the metric equation becomes
\begin{equation}
\mathcal E^{(0)}_{\mu\nu}+\bar\Lambda g_{\mu\nu}
-\kappa_\Lambda\left(
\nabla_\mu\Lambda\nabla_\nu\Lambda
-\frac12g_{\mu\nu}(\nabla\Lambda)^2
-\frac12m_\Lambda^2(\Lambda-\bar\Lambda)^2g_{\mu\nu}
\right)=0 .
\end{equation}
This is not the initial gravity with a locally varying cosmological constant in the geometric sector.  It is the initial theory with constant cosmological constant \(\bar\Lambda\), plus the stress tensor of a massive scalar fluctuation \(\Lambda(x)-\bar\Lambda\).  This distinction prevents one from interpreting the construction directly as a quintessence model.  A genuine dark-energy model would require a separate FRW analysis, including the equation of state and the allowed scalar mass scale.

The quadratic potential resembles the potential generated after integrating out non-propagating form fields in Kaloper-Sorbo-type constructions \cite{Kaloper:2008fb,Kaloper:2011jz,Kaloper:2014dqa}.  The analogy is structural, not phenomenological.  Here the scalar relaxes to the charge value \(\bar\Lambda\), and the thermodynamic coupling charge is the integration constant \(\bar\Lambda\), not the local scalar value.

\subsection{The Gauss-Bonnet coupling}

The reason for using the defect is clearer for a genuinely spacetime-dependent density.  In \(D\ge5\), take
\begin{equation}
        \mathcal L=\mathcal L_0-\bar\alpha_{\rm GB}\mathcal L_{\rm GB},
        \qquad
        \mathcal L_{\rm GB}=\mathcal G
        =R^2-4R_{\mu\nu}R^{\mu\nu}
        +R_{\mu\nu\rho\sigma}R^{\mu\nu\rho\sigma} .
\end{equation}
Promote \(\bar\alpha_{\rm GB}\) to a scalar-gauge pair and define
\begin{equation}
        \Delta_{\rm GB}=F_{\rm GB}-\mathcal G .
\end{equation}
On the constant-coupling sector, \(\alpha_{\rm GB}=\bar\alpha_{\rm GB}\) and \(\Delta_{\rm GB}=0\), so the top-form component tracks the curvature density:
\begin{equation}
        F_{\rm GB}=\mathcal G(x) .
\end{equation}
This is precisely what a bare \(F_{\rm GB}^2\) kinetic term would obstruct.

For pure constant-curvature AdS, with
\begin{equation}
R_{\mu\nu\rho\sigma}
=-\frac{1}{\ell^2}
(g_{\mu\rho}g_{\nu\sigma}-g_{\mu\sigma}g_{\nu\rho}),
\end{equation}
one has 
\begin{equation}\label{GB-AdS-correct}
        \mathcal G_{\rm AdS}
        =\frac{D(D-1)(D-2)(D-3)}{\ell^4} .
\end{equation}
On an AdS black hole background, the curvature density contains this asymptotic part plus a generally non-constant black hole contribution.  Schematically,
\begin{equation}
        \mathcal G(r)=\mathcal G_{\rm AdS}+\mathcal G_{\rm BH}(r),
        \qquad
        \mathcal G_{\rm BH}(r)\to0
        \quad\text{as}\quad r\to\infty .
\end{equation}
The defect construction allows \(F_{\rm GB}\) to follow this non-constant density while keeping \(\Delta_{\rm GB}=0\) and no additional backreaction from the completion in the constant-coupling sector.

When the completion is excited, the deviation obeys
\begin{equation}
        (\Box-m_{\rm GB}^2)(\alpha_{\rm GB}-\bar\alpha_{\rm GB})=0,
        \qquad
        m_{\rm GB}^2=\frac{1}{\kappa_{\rm GB}\hat\kappa_{\rm GB}} .
\end{equation}
The asymptotic charge remains \(\bar\alpha_{\rm GB}\), while the local scalar excitation contributes only through its stress tensor.  Thus, the Gauss-Bonnet example illustrates the basic generality of the defect construction: the defect can accommodate a non-constant density \(\mathcal L_i(x)\) without spoiling the constant-coupling theory, and its excitation gives a localized massive-scalar backreaction rather than a local replacement of the original coupling by \(\alpha_i(x)\).
}

\section{Conclusions}

We have constructed a dynamical completion of the scalar-gauge pair formulation in which coupling constants appear as conserved charges in black hole thermodynamics.  The key ingredient is the shifted defect \(\boldsymbol\Delta_i=\boldsymbol{\mathrm{F}}_i-\mathbf L_i\).  Kinetic terms built from this defect vanish on the original constraint surface, so all constant-coupling black hole solutions and their extended thermodynamic relations remain intact.

The main lesson is also a limitation.  Once the top-form is eliminated, the original geometric and matter equations are governed by the constant integration parameter \(\bar\alpha_i\), not by the local field \(\alpha_i(x)\).  The local deviation \(\alpha_i(x)-\bar\alpha_i\) is a new massive scalar degree of freedom with positive energy for \(\kappa_i>0\) and \(\hat\kappa_i>0\). 

This conservative interpretation is the one directly supported by the field equations. It is also consistent with the covariant phase-space charge calculation: the charge associated with the global part of the form-field gauge symmetry is \(\bar\alpha_i\).  It also explains why the construction works for general densities, such as the Gauss-Bonnet invariant.  The top-form component may track a non-constant \(\mathcal L_i(x)\), while the defect vanishes on the constant-coupling solution.

The cosmological example should be read with the same care.  The framework gives Einstein gravity with constant \(\bar\Lambda\) plus a massive scalar stress tensor, not automatically a viable quintessence or dark-energy model.  Such an application would require solving the cosmological equations and deriving the equation of state.  Likewise, stationary black hole solutions with nonzero scalar excitation require a separate existence and no-hair analysis.  These are natural directions for future work, but they are not assumptions made in the present construction. 

\appendix
\section{Asymptotic conditions}
Let us explain \eqref{falloff-flat} and \eqref{falloff-ads}. Let 
       \( \phi_i:=\alpha_i-\bar\alpha_i \).
In the asymptotic region, the scalar equation reduces to
\begin{equation}
        (\Box-m_i^2)\phi_i=0 .
\end{equation}
For a static, spherically symmetric asymptotic metric of the form
\begin{equation}
        \mrd s^2=-f(r)\,\mrd t^2+\frac{\mrd r^2}{f(r)}+r^2\mrd\Omega_{D-2}^2 ,
\end{equation}
the radial equation is
\begin{equation}
        \frac{1}{r^{D-2}}\frac{\mrd}{\mrd r}
        \left(
        r^{D-2}f(r)\frac{\mrd\phi_i}{\mrd r}
        \right)
        -m_i^2\phi_i=0 .
\end{equation}
This radial equation gives the two standard asymptotic falloffs.  In an
asymptotically flat region \(f(r)\to1\), and therefore
\begin{equation}
      \phi_i''+\frac{D-2}{r}\phi_i'-m_i^2\phi_i=0,
        \qquad
        \phi_i(r)\sim \frac{e^{-m_i r}}{r^{(D-2)/2}} .
\end{equation}
In an asymptotically AdS region \(f(r)\sim r^2/\ell^2\), and therefore
\begin{equation}
        \phi_i''+\frac{D}{r}\phi_i'
        -\frac{m_i^2\ell^2}{r^2}\phi_i=0 .
\end{equation}
The ansatz \(\phi_i\sim r^{-\delta}\) gives
\begin{equation}
        \delta(\delta-D+1)=m_i^2\ell^2,
\end{equation}
so that
\begin{equation}
        \delta_\pm
        =
        \frac{D-1}{2}
        \pm
        \sqrt{
        \frac{(D-1)^2}{4}+m_i^2\ell^2
        } .
\end{equation}
The normalizable AdS branch is \(\phi_i\sim r^{-\delta_+}\).

\end{document}